\documentclass[10pt]{article} 	% use \documentstyle for old LaTeX compilers

\usepackage[ngerman]{babel} 	% 'french', 'german', 'spanish', 'danish', etc.
\usepackage{amssymb}
\usepackage{comment}
\usepackage{amsmath}
\usepackage{txfonts}
\usepackage{mathdots}
\usepackage[classicReIm]{kpfonts}
\usepackage{lmodern}
\usepackage{graphicx}
\usepackage{xcolor}
\usepackage{setspace}
\usepackage{times}
\usepackage{fancyhdr} 
\usepackage{scrextend}
\usepackage[colorlinks=false,hidelinks]{hyperref}
\usepackage{enumitem}
\usepackage{indentfirst}
\usepackage{tabularx, multirow}
\usepackage{float, csquotes}
\newcolumntype{Y}{>{\centering\arraybackslash}X} %tabuarlx, centered
\newcolumntype{Z}{>{\raggedleft\arraybackslash}X} %tabularx, right alignment
\newcolumntype{A}[1]{>{\raggedright\arraybackslash}p{#1}} %tabularx, left alignment
\newcolumntype{C}[1]{>{\centering\arraybackslash}p{#1}}
\usepackage{todonotes}
\setuptodonotes{color=blue!30}
\usepackage[T1]{fontenc}
\usepackage[printonlyused, nohyperlinks]{acronym}
\acrodef{milp}[MILP]{gemischt ganzzahlige lineare Optimierung}
\acrodef{eom}[EOM]{Energie-Options-Modell}
\acrodef{rmse}[RMSE]{Wurzel des mittleren quadratischen Fehlers}
\acrodef{nrmse}[nRMSE]{normalisierte Wurzel des mittleren quadratischen Fehlers}
\addto\extrasngerman{}
\addto\extrasngerman{}
\addto\extrasngerman{}
\usepackage[style=ieee,sorting=none,uniquename=init,defernumbers=true, maxcitenames=1, minbibnames = 6, maxbibnames=6, mincitenames=1, 
doi=true,url=true,isbn=false
]{biblatex}
\DefineBibliographyStrings{ngerman}{			% Ändert "u.a." in "et al." bei \citeauthor
	andothers = {{\textit{et\,al\adddot}}},
}
\newcommand{\citet}[1]{\citeauthor{#1}~\cite{#1}}
\usepackage{etoolbox}
\makeatletter
\patchcmd{\NAT@test}{\else\NAT@nm}{\else\NAT@nmfmt{\NAT@nm}}{}{}
\let\NAT@up\itshape
\makeatother
\usepackage{xcolor}
\definecolor{hsurot}{cmyk}{.00 1 .59 .26}
\definecolor{hsugrau}{cmyk}{.38 .37 .39 .15}
\definecolor{hsublau}{cmyk}{1 .40 0 .82}
\definecolor{hsugelb}{cmyk}{0 .16 .80 0}
\definecolor{hsuturkis}{cmyk}{1 .14 .60 .49}
\definecolor{hsugruen}{cmyk}{.16 .16 .91 .28}
\definecolor{hsubraun}{cmyk}{.00 .57 1 .17}
\definecolor{hsuorange}{cmyk}{.01 .87 .77 .13}
\usepackage{tikz}
\usetikzlibrary{positioning,chains,external,shapes.geometric, spy, external}
\usepackage{pgfplots}
\usepackage{pgfplotstable} % also loads pgfplots
\usepgfplotslibrary{patchplots, colormaps, groupplots, dateplot}
\pgfplotsset{compat=newest}
\usepackage{subcaption}
\usepackage{booktabs}

\usepackage[
	left=2.5cm,
	right=2.5cm,
	top=2.5cm,
	bottom=2cm,
]{geometry}

\fancypagestyle{ErsteSeite}{%
	\fancyhf{}%
	\fancyhead[L]{\color{gray} Wissenschaftlicher Hauptbeitrag atp magazin {\textbar} Formatvorlage 2020}
}
\begin{document}
%\selectlanguage{english} % remove comment delimiter ('%') and select language if required

% Kopfzeile
\thispagestyle{ErsteSeite}

% Titel und Subtitel
\color{black}
\begin{addmargin}[0cm]{3cm}
	{\fontfamily{phv}\selectfont
		\fontsize{28pt}{12pt}\selectfont
		\noindent \textbf{Nutzung von Massespeichern zur Flexibilisierung des Energieverbrauchs} \vspace{14,2pt}
		
		\fontsize{20pt}{12pt}\selectfont
		\noindent \textbf{Kosteneffizienter Anlagenbetrieb durch Anpassung an Marktpreise} \vspace{14.2pt}
	}
\end{addmargin}
\pagestyle{empty} % Kopf- und Fußzeilen auf allen weiteren Seiten entfernen

% Autoren
\fontsize{10pt}{12pt}\selectfont
\begin{onehalfspace}
	\noindent Lukas Peter Wagner, Lasse Matthias Reinpold, Maximilian Kilthau, Felix Gehlhoff,\\ Institut für Automatisierungstechnik, Helmut-Schmidt Universität, Hamburg\vspace{6pt}
 
    \noindent Christian Derksen, Nils Loose,  EnFlexIT GmbH, Essen \vspace{6pt}

 \noindent Julian Jepsen, Institut für angewandte Werkstofftechnik,  Helmut-Schmidt Universität, Hamburg und \\ Institut für Wasserstofftechnologie,  Helmholtz-Zentrum Hereon, Geesthacht\vspace{6pt}
		
	\noindent Alexander Fay, Lehrstuhl für Automatisierungstechnik, Ruhr Universität, Bochum\vspace{6pt}
\end{onehalfspace}

% Kurzfassung Deutsch
\begin{addmargin}[0cm]{5cm}
\noindent Der zunehmende Anteil erneuerbarer Energien und die damit verbundene Notwendigkeit, die Energieerzeugung und den Energieverbrauch an die Verfügbarkeit von Energie anzupassen, erfordern neue Konzepte zur energieflexiblen Steuerung industrieller Produktionsanlagen. In diesem Beitrag wird daher demonstriert, wie das Potenzial von Massespeichern zur Steigerung der Energieflexibilität in industriellen Prozessen, insbesondere durch die Anwendung optimierter Betriebsplanung basierend auf Marktpreisen, genutzt werden kann. Als Beispiel wird eine Abwasseraufbereitungsanlage mit Dekantern und zugehörigen Massespeichersystemen betrachtet, um deren Energieflexibilität optimiert zu nutzen. 
Ein gemischt-ganzzahliges, lineares Optimierungsmodell wurde entwickelt und in einem realen Feldversuch getestet, um die Energiekosten des Betriebs der Abwasseraufbereitungsanlage zu minimieren. Das Optimierungsmodell wurde in einer eigens entwickelten Optimierungsumgebung für energieflexible Produktionsanlagen erstellt. Der resultierende Betriebsplan wurde in einem ebenfalls eigens entwickelten Assistenzsystem an die Anlagenbetreiber übermittelt und über einen Zeitraum von über 24 Stunden ausgeführt. Das Modell zeigt eine gute Übereinstimmung mit den realen Messwerten, mit durchschnittlichen Fehlern zwischen 3~\% und 7~\%. Alle von den Betreibern gesetzten Betriebsziele wurden erreicht, was die Anwendbarkeit des Modells in industriellen Umgebungen bestätigt. Die Ergebnisse zeigen ein erhebliches Einsparpotenzial von etwa 56~\% für den untersuchten Zeitraum. 
	
\textit{} \vspace{12pt}

\noindent Energieflexibilität, Optimierung, Betriebsplanung,  Energiekosten,  Massespeicher, Abwasseraufbereitungsanlage \vspace{20pt}

% Titel auf Englisch
\fontsize{20pt}{12pt}\selectfont
{\fontfamily{phv}\selectfont
	\noindent \textbf{Utilizing Mass Storage for Flexibilizing Energy Resource Operation} \\
 {\normalsize Cost-Efficient Resource Operation by Responding to Market Prices\textit{}}\vspace{14.2pt}
}

% Kurzfassung Englisch
\fontsize{10pt}{12pt}\selectfont
	\noindent The increasing share of renewable energy sources and the associated need to align energy production and consumption with energy availability necessitate new concepts for energy flexible operation of industrial production resources. In this paper, we demonstrate the potential of mass storage to increase energy flexibility in industrial operations, specifically through the application of optimized operational planning based on market prices. 
 As a case study, a wastewater treatment plant equipped with decanters and associated storage systems is examined to optimally utilize its Energy-flexibility. A mixed-integer linear programming optimization model was developed and tested in a real-world field trial to minimize the power cost of the wastewater treatment plant's operation. The optimization model was created in a specially developed optimization environment for energy-flexible production facilities. The resulting operation plan was transmitted to the plant operators via a specially developed assistance system and executed over a period of more than 24 hours. The model shows good agreement with real measurements, exhibiting average errors between 3~\% to 7~\%. All operational goals set by the operators were fulfilled, validating the model's applicability in industrial settings. The results demonstrate significant potential for cost savings of roughly 56~\% for the investigated time horizon. 
  \vspace{12pt}

\noindent Energy Flexibility / Optimization / Operational Planning / Energy Cost / Mass Storage / Wastewater Treatment Plant

% \pagebreak
\section{Einleitung} \label{sec:intro}
Der zunehmende Anteil erneuerbarer Energien und die damit verbundene Notwendigkeit der Anpassung der Energieerzeugung und des -verbrauchs an die Verfügbarkeit von Energie erfordern neue Konzepte zur energieflexiblen Steuerung von Anlagen durch optimierte Fahrpläne, welche den Energieverbrauch oder die -erzeugung  planen \cite{KBU+24}. 

Ein typischer Preisverlauf des Intraday-Markts, wie in \autoref{fig:strompreis-epex-feldtest} dargestellt, unterliegt Schwankungen, die vor allem durch die erhöhte Energieerzeugung durch Photovoltaik bei hoher Sonneneinstrahlung und Windturbinen in Zeiten hoher Windgeschwindigkeiten beeinflusst werden. Dadurch spiegeln die viertelstündlichen Preissignale gut das Angebot von Energie wider und ermöglichen die preisbasierte Anpassung des Energieverbrauchs, welche großes finanzielles Einsparpotenzial birgt~\cite{RWK+23}.

\begin{figure}[h]
    \centering
    \includegraphics[width=.75\linewidth]{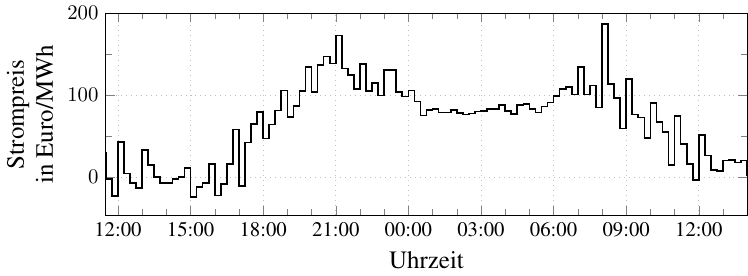}
   % \resizebox{.75\linewidth}{!}{\input{Abbildungen/elprice}}
    \caption{Dynamischer Strompreis des kontinuierlichen Intraday-Markts \cite{EPEb}}
    \label{fig:strompreis-epex-feldtest}
\end{figure}

% \todo[inline]{Energiespeicherstrategie \cite{Bun23}}
Die Theorie der Planung des energieflexiblen Betriebs von Anlagen ist bereits gut erforscht: durch Modellierung der Anlagen zur Optimierung können optimierte Betriebspläne berechnet werden \cite{WRK+23}. Diese können dann in Sollwerte zur Steuerung der Anlagen gewandelt werden \cite{RWF23}. Insbesondere der Zusammenhang zwischen Energieverbrauch und Produktionsmenge o.Ä. spielt in Optimierungsmodellen eine zentrale Rolle \cite{WRF23b}, da nur so sichergestellt werden kann, dass Energiekosten verringert und gleichzeitig Produktionsziele eingehalten werden.

Die Analyse bestehender Arbeiten, welche die Umsetzung optimierter Betriebspläne beschreiben,  zeigt jedoch, dass, obwohl Modelle und Ansätze zur Optimierung des Energieverbrauchs einer Vielzahl von Anlagen präsentiert werden \cite{WRK+23}, es nur eine begrenzte Anzahl an Arbeiten gibt, die diese Modelle unter realen Betriebsbedingungen validieren. Dies jedoch nur in kontrollierten Laborumgebungen: \citet{BHA+23} untersuchen die  Optimierung des Betriebs von Wärmepumpen. \citet{KSB+24} entwickeln einen Optimierungsansatz zur Abwärmerückgewinnung in der Stahlproduktion. \citet{PCF+19} beschreiben einen experimentellen Aufbau für den optimierten Betrieb von Wärmepumpen mit variabler Leistung, um die Energieflexibilität von Gebäuden zu erhöhen.
Ebenfalls wird oft nicht der Einfluss der Veränderung des Energieverbrauchs auf die Produktionsmenge der Anlage betrachtet \cite{GFS+22}. Die praktische Implementierung und Auswirkungen auf den laufenden Betrieb in industriellen Umgebungen bleiben daher oft ungetestet.

Im Projekt \textit{OptiFlex} wird untersucht, wie energieverbrauchende und -erzeugende Anlagen energieflexibel betrieben werden können und welche Potenziale sich durch eine solche Betriebsweise realisieren lassen. Ebenfalls wird die Umsetzung energieflexibler Betriebsweisen in der realen Anwendung getestet.

Dieser Beitrag untersucht die Forschungsfrage,  welches Energieflexibilitätspotenzial eine Abwasseraufbereitungsanlage besitzt, die aus Dekantern sowie einem vorgelagerten Speicherbecken (Schlammtasche) für zu entwässernden Dünnschlamm und nachgelagerten Containern für entwässerten Trockenschlamm besteht. Das Flexibilitätspotenzial wird genutzt, indem der Betrieb der Anlage zu Zeiten geringer Strompreise an der Intraday-Strombörse geplant wird.

In \autoref{sec:method} wird auf Basis eines generischen Optimierungsmodells ein Modell für den zu untersuchenden Anlagenverbund der Abwasseraufbereitungsanlage erstellt und parametriert. Dies geschieht in einer eigens für die Bearbeitung dieses Forschungsthemas geschaffenen Optimierungsumgebung. In \autoref{sec:ergebnisse} wird die Anwendung des Modells zur Generierung eines optimierten Betriebsplans beschrieben. Der optimierte Fahrplan wird durch den realen Anlagenverbund umgesetzt. Der sich hieraus ergebende Mehrwert wird beschrieben. Darauf folgend werden das methodische Vorgehen und die Ergebnisse in \autoref{4disconcl} zunächst diskutiert und dann ein Fazit gezogen.

\section{Ansatz zur Flexibilisierung des Anlagenbetriebs} \label{sec:method}
\noindent In diesem Abschnitt wird der Ansatz zur Flexibilisierung des Anlagenbetriebs durch Optimierung vorgestellt. In \autoref{sec:anlagenverbund} wird der zu flexibilisierende Anlagenverbund beschrieben. Folgend wird in \autoref{sec:method:modell} das Optimierungsmodell zur Abbildung des Verhaltens einzelner Anlagen innerhalb eines Verbunds beschrieben, welches mit den in \autoref{sec:method:parameter} beschriebenen Parametern parametriert wird. Die softwaretechnische Umsetzung des Modells sowie die Anwenderführung wird in \autoref{sec:method:eom} erläutert.

\subsection{Beschreibung des Anlagenverbunds} \label{sec:anlagenverbund}
Die in \autoref{fig:aufbau:decanter} dargestellte Abwasseraufbereitungsanlage wird im Rahmen des Feldtests energieflexibel betrieben. Die Anlage besteht aus einer Schlammtasche, zwei Dekantern und  vier Containern. 

\begin{figure}[h]
    \centering
    \includegraphics[width=10cm]{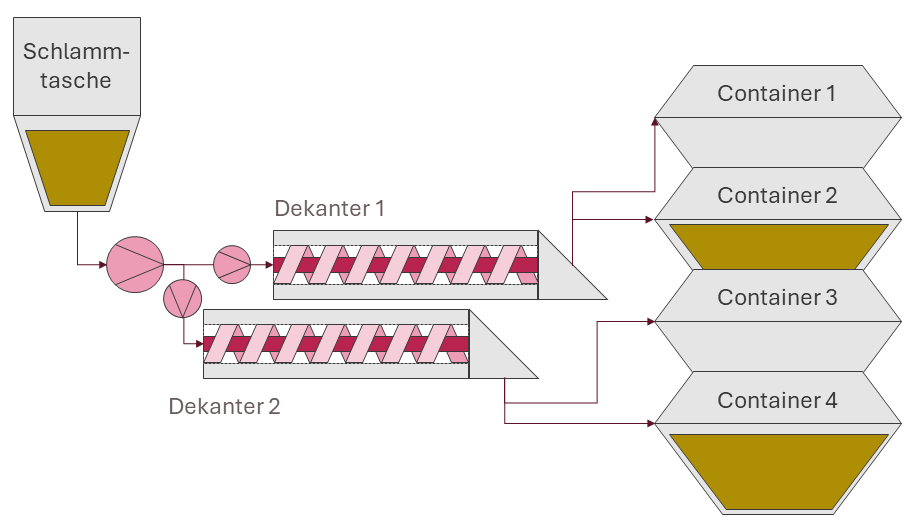}
    \caption{Aufbau der Abwasseraufbereitungsanlage}
    \label{fig:aufbau:decanter}
\end{figure}

Die Schlammtasche wird durch das einfließende Abwasser aus umliegenden Industriebetrieben befüllt. Der sich absetzende \textit{Dünnschlamm} wird in zwei Dekantern, unter Nutzung elektrischer Energie, entwässert und der \textit{Trockenschlamm} in jeweils zwei Container pro Dekanter gefördert. Die Container werden dann zu vordefinierten Zeitpunkten ausgetauscht, wenn \textit{genug} Trockenschlamm enthalten ist.

Die Dekanter stellen in diesem Anlagenverbund die steuerbaren flexiblen Anlagen dar, während die Schlammtasche und die Container als Massespeicher fungieren. Durch die Kombination der flexibel steuerbaren Dekanter mit den vor- und nachgelagerten Massespeicher entsteht das Flexibilitätspotenzial des Anlagenverbunds \cite{ZKZ+21}.

Das die Anlage betreibende Unternehmen agiert bereits am Intraday-Markt, da unternehmenseigene elektrische Energieerzeugung vermarktet wird. Ein energie\-flexibler Betrieb der Abwasseraufbereitungsanlage findet bis dato jedoch nicht statt.  
\subsection{Optimierungsmodell zur Planung des Anlagenbetriebs} \label{sec:method:modell}
Eine Analyse bestehender generischer gemischt ganzzahliger linearer (\acsu{milp}) Optimierungsmodelle durch \citet{WRF23b} ergab, dass zur Beschreibung des Verhaltens energetisch flexibler Anlagen Nebenbedingungen für Betriebsgrenzen, für den Zusammenhang zwischen Ein- und Ausgangsgrößen und für Systemzustände notwendig sind. Darüber hinaus werden Energiespeicher über eine Energiebilanz abgebildet, und Massespeicher wie die in \autoref{sec:intro} beschriebene Schlammtasche und Container, werden über eine Massebilanz abgebildet. Zur Aggregation mehrerer Anlagen innerhalb eines Verbunds müssen zudem die Verbindungen zwischen einzelnen Anlagen abgebildet werden.

Die mathematische Formulierung der im Rahmen des Feldtests genutzten Nebenbedingungen ist im Folgenden dargestellt und basiert auf der von \citet{WRF23b} entwickelten Modellstruktur. 

Die Betriebsgrenzen der einzelnen Dekanter werden, wie in    \autoref{eq:betriebsgrenzen} beschrieben, über einen Wertebereich für die jeweilige Entscheidungsvariable des Betriebspunks OP modelliert. 
\begin{align}
\text{OP}_{\text{min}} \le \text{OP}_t \le \text{OP}_{\text{max}} \label{eq:betriebsgrenzen}
\end{align}

Leistungen und Materialflüsse werden mit $P$ gekennzeichnet. Der Zusammenhang zwischen elektrischer Leistungsaufnahme $P_{\text{el}}$ sowie Dünnschlamm- $P_{\text{DS}}$  und Trockenschlamm-Durchsatz $P_{\text{TS}}$  wird durch den Betriebspunkt mithilfe von Parametern \textit{a,b,c,d,e}  abgebildet (siehe \autoref{eq:Leistungsaufnahme} bzw.  \autoref{eq:Duennschlamm-Durchsatz}). 

\begin{align}
 P_{\text{el},t}=\; &a \cdot x_{\texttt{Betrieb}, t} \label{eq:Leistungsaufnahme} \\
+&b \cdot \text{OP}_{t} \notag \\
+&c \cdot \text{Trockenschlammdichte}_t \cdot \notag x_{\texttt{Betrieb},t} \notag \\
+&d \cdot x_{\texttt{Start}, t}   \notag \\
 P_{\text{DS},t}=\; & e \cdot \text{OP}_{t}    \label{eq:Duennschlamm-Durchsatz}
 \end{align}
 
Der Trockenschlamm-Durchsatz wird, wie in \autoref{eq:trockenschlamm} dargestellt, analog zur Softsensor-basierten Messung im Prozessleitsystem durch Multiplikation des Dünn\-schlamm-Volumenstroms mit der Trockenschlamm-Dichte ermittelt.
 \begin{align}
 P_{\text{TS},t}= P_{\text{DS},t} \cdot \text{Trockenschlammdichte}_t \label{eq:trockenschlamm}
 \end{align}

Die für den energetisch flexiblen Betrieb des Anlagenverbunds relevante elektrische Leistung ist die Summe der elektrischen Eingangsleistung beider Dekanter (siehe \autoref{eq:psystem}). 
\begin{align}
    P_{\text{System},t} =  P_{\text{el, Dekanter 1},t} + P_{\text{el, Dekanter 2},t} \label{eq:psystem}
\end{align}

Binäre Variablen $x_s$ werden zur Abbildung einzelner Systemzustände $s$ benötigt ($s \in \mathcal{S}$). Die notwendigen Nebenbedingungen zur Abbildung der Systemzustände beinhalten die Wahl des jeweils aktiven Zustands (Gleichungen~\ref{eq:state:lb}--\ref{eq:stateeq1}), definiert durch untere ($\text{OP}_{\text{min}, s}$) und obere ($\text{OP}_{\text{max}, s}$) Grenzen des Betriebspunkts je Zustand $s$. Es kann jeweils nur ein Zustand je Zeitschritt $t$ aktiv sein (siehe \autoref{eq:stateeq1}.

 \begin{align}
     \text{OP}_{t} &\geq \sum \limits_{s \in \mathcal{S}}  \text{OP}_{\text{min}, s} \cdot x_{s, t} \label{eq:state:lb}\\
     \text{OP}_{t} &\leq \sum \limits_{s \in \mathcal{S}}  \text{OP}_{\text{max}, s} \cdot x_{s,t} \label{eqeq:state:ub} \\
     &\sum \limits_{s \in \mathcal{S}} x_{s, t} = 1 \label{eq:stateeq1} 
     \end{align}

Des Weiteren sind noch Gleichungen zur Auswahl eines Folgezustands des jeweils aktiven Systemzustands aus der Menge der möglichen Folgezustände $\mathcal{S}_{F,s}$ (\autoref{eq:Folgezustände}) und zur Einhaltung der sog. Haltedauern: der minimalen und maximalen Aktivität eines Systemzustands ($t_{h, \text{min},s}$ bzw. $t_{h, \text{max},s}$, Gleichungen~\ref{eq:haltedauermin} und~\ref{eq:haltedauermax}). 
     \begin{align}
 x_{t-1,s} - x_{t,s} &\le \sum \limits_{f \in \mathcal{S}_{F,s}} x_{f,t} \label{eq:Folgezustände} \\
    t_{h, \text{min},s} \cdot  \left( x_{t,s} - x_{t-1,s}   \right) &\le \sum \limits_{\tau \in \mathcal{T}_h} x_{\tau,s} \label{eq:haltedauermin}  \\
    t_{h, \text{max},s} &\ge   \sum \limits_{\tau \in \mathcal{T}_h} x_{\tau,s} \label{eq:haltedauermax}
 \end{align}

Zusätzlich ist eine Nebenbedingung enthalten, die Lastspitzen beim Anfahren verhindert (\autoref{eq:lastspitze}), indem sie definiert, dass sich nicht beide Dekanter gleichzeitig im \texttt{Start}-Zustand befinden dürfen. Dies wurde gemacht, da im \texttt{Start}-Zustand der Dekanter Spitzenlasten auftreten. Die Möglichkeit des gleichzeitigen Betriebs beider Dekanter ist betrieblich notwendig, daher wird eine analoge Nebenbedingung für den Zustand \texttt{Betrieb} beider Dekanter nicht benötigt. 

\begin{align}
 x_{\text{\texttt{Start}, Dekanter 1},t} +x_{\text{\texttt{Start}, Dekanter 2},t} \leq 1 \label{eq:lastspitze}
 \end{align}

Die in \autoref{eq:soc2} dargestellte Energiebilanz für Speichersysteme ist entsprechend vereinfacht, da im Rahmen des Feldtests nur Materialspeicher mit einem Wirkungsgrad von 100~\% betrachtet werden. \autoref{eq:sco:bounds} begrenzt den Füllstand, sodass die Betriebsgrenzen des Speichers eingehalten werden.
\begin{align}
  \text{SOC}_{t} = \text{SOC}_{t-1}  + \left(P_{\text{in}, t} - P_{\text{out}, t}\right) \cdot \Delta t \label{eq:soc2} \\
  \text{SOC}_{\text{min}} \leq \text{SOC}_t \leq \text{SOC}_{\text{max}} \label{eq:sco:bounds}
  \end{align}

Der Optimierungshorizont $\mathcal{T}$ ist in äquidistante Zeitschritte $t$ mit Länge $\Delta t$ eingeteilt. 

Die dargestellten Nebenbedingungen werden in Kombination mit der in \autoref{eq:obj} formulierten Zielfunktion in einem Optimierungsmodell für die Abwasseraufbereitungsanlage genutzt. Die Zielfunktion strebt eine Minimierung der Energiekosten im  Optimierungshorizont (=Zeitraum des Feldtests) unter Berücksichtigung der jeweiligen Strompreise  $k_{\text{el},t}$ an (\autoref{fig:strompreis-epex-feldtest}). Im Kontext des optimierten Betriebs von Anlagen sind auch andere Zielfunktionen denkbar, wie beispielsweise die Minimierung von CO$_2$-Emissionen \cite{WRK+23, RWK+23}.
\begin{align}
    \min \text{Kosten} = \sum \limits_{t\in \mathcal{T}} P_{\text{System},t} \cdot \Delta t \cdot k_{\text{el},t}\label{eq:obj}
\end{align}

\subsection{Parametrierung des Optimierungsmodells} \label{sec:method:parameter}
In \autoref{tab:parameter} werden die Parameter zur Nutzung des in \autoref{sec:method:modell} beschriebenen Modells dargestellt. Die dort gelisteten Parameter werden jeweils für beide Dekanter genutzt. Dort enthalten sind Parameter für Betriebsgrenzen, leistungsbezogene Parameter, Systemzustände, mögliche Folgezustände (darstellt als mögliche Zustandsübergänge), sowie Haltedauern der jeweiligen Zustände. 

Weitere Betriebsgrenzen sind nicht notwendig, da der Betriebspunkt \textit{OP} bereits alle Zusammenhänge begrenzend beeinflusst. Zusätzlich wurde noch ein Wert mit erwartetem Dünnschlammzustrom in die Schlammtasche von 10~m$^3$/h und ein Wert für die erwartete Trockenschlammdichte von 30~$g/l$ im Modell hinterlegt. 

\begin{table}[h]
    \centering
    \caption{Parameter des Optimierungsmodells der Dekanter}
    \label{tab:parameter}
\begin{tabularx}{.75\linewidth}{A{.25\linewidth}C{.5\linewidth}}
\toprule
\textbf{Beschreibung}        &  \textbf{Parameter} \\
\midrule
Betriebsgrenzen OP & min. 0, max. 1 \\
\midrule
\multirow{2}{*}{Leistungsbezogene Parameter} & 
 a = 22,356 kW \\
 & b = 8,464 kW \\
 &  c = 0,0182 (kW$\cdot$l)/g\\
 &d = 21,366 kW\\
 & e=  12 m$^3$/h \\
\midrule
Zustände & \texttt{Aus},  \texttt{Start}, \texttt{Betrieb}\\
\midrule
Zustandsübergänge & \texttt{Aus} $\rightarrow$ \texttt{Start},\\
& \texttt{Start}  $\rightarrow$ \texttt{Betrieb}, \\
& \texttt{Betrieb} $\rightarrow$ \texttt{Aus} \\
\midrule
Haltedauern \texttt{Aus} & min: 60 min.,  max: $\infty$ \\
Haltedauern \texttt{Start} & min: 6 min., max: 6 min.  \\
Haltedauern \texttt{Betrieb} & min: 60 min., max:  $\infty$ \\
\bottomrule
\end{tabularx}
\end{table}

\subsection{Anwenderführung zur Umsetzung des energieflexiblen Betriebsplans} \label{sec:method:eom}
Zur Umsetzung des Optimierungsmodells und nachfolgenden Steuerung des Anlagenverbunds auf Basis eines energieflexiblen Betriebsplans wird das \ac{eom} genutzt. Das \ac{eom} ist eine Modellierungsumgebung, mit der Energiesysteme auf Basis einer Zustandsbeschreibung modelliert und nachfolgend für Simulationen oder Optimierungen genutzt werden können \cite{DeUn16}. Die Art der Bewertung ist frei konfigurierbar \cite{DeUn16} und wurde im Projekt \glqq OptiFlex \grqq \ durch die Möglichkeit zur \ac{milp}-Optimierung erweitert. Wie in \autoref{fig:optimierungsumgebung} dargestellt, kann im Bereich der Bewertung das Optimierungsproblem, bestehend aus Entscheidungsvariablen, Parametern (in \autoref{fig:optimierungsumgebung} zusammengefasst unter 'Optimization Variables'), Nebenbedingungen ('Constraint Functions') und Zielfunktion ('Objective Function'), definiert werden. 

\begin{figure}[h]
    \centering
        \includegraphics[width = .6\linewidth]{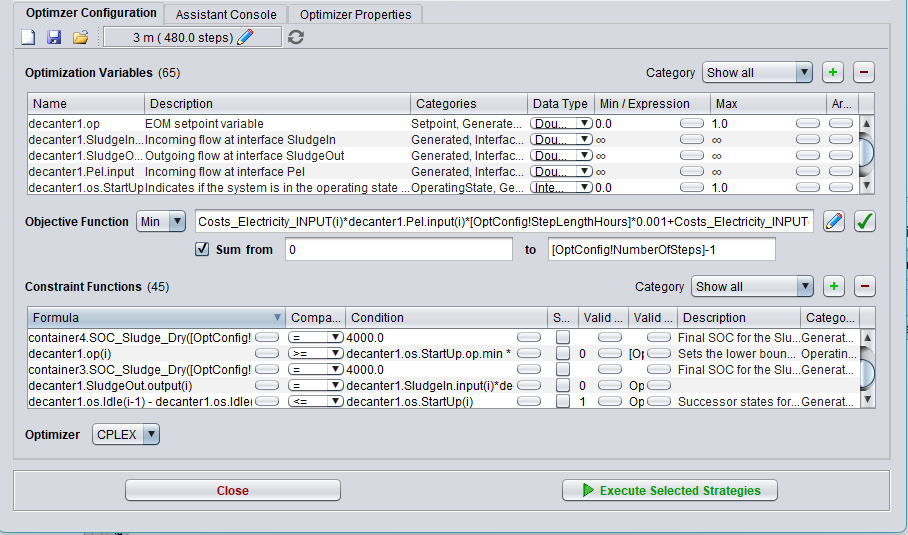}
    \caption{Optimierungsumgebung im \ac{eom}}
    \label{fig:optimierungsumgebung}
\end{figure}

Alle Elemente des Optimierungsmodells können aus der Definition des \ac{eom}-Modells generiert werden. Die Modellierung folgt der in \autoref{sec:method:modell} beschriebenen Form. So können beispielsweise Entscheidungsvariablen und Nebenbedingungen für Systemzustände, Zustandsfolgen und Haltedauern automatisch generiert werden. Die Form, in der das Optimierungsproblem gespeichert wird, ist unabhängig vom verwendeten Solver. Die im Optimierungsproblem enthaltenen Entscheidungsvariablen, Parameter, Nebenbedingungen und die Zielfunktion werden in eine den jeweiligen Solver geeignete Form gewandelt. Im Rahmen der Anwendung im Projekt kommt IBM ILOG CPLEX als Solver zum Einsatz. Die Einbindung anderer Solver ist ebenfalls möglich, erfordert jedoch die Anpassung der Modell-Programmierung. IBM ILOG CPLEX wurde ausgewählt, da dieser Solver sehr leistungsstark sowie vergleichbar mit Gurobi ist und durch die Schnittstelle zu Java eine gute Integration in die bestehende Software möglich ist.

Nach dem Lösen des Optimierungsmodells kann der generierte Betriebsplan entweder direkt an das Prozessleitsystem der zu steuernden Anlage(n) übermittelt werden (bspw. per OPC-UA, wie von \citet{RWF23} demonstriert) oder als Handlungsempfehlung zu bestimmten Zeitpunkten für den Anlagenbetreiber angezeigt werden. \autoref{fig:eom:assi} zeigt das Assistenzsystem, in dem Handlungsempfehlungen im Sinne von einzustellenden Werten an der Anlage zu bestimmten Zeitpunkten innerhalb des Optimierungshorizonts angezeigt werden.

\begin{figure}[H]
    \centering
    \includegraphics[width=0.75\linewidth]{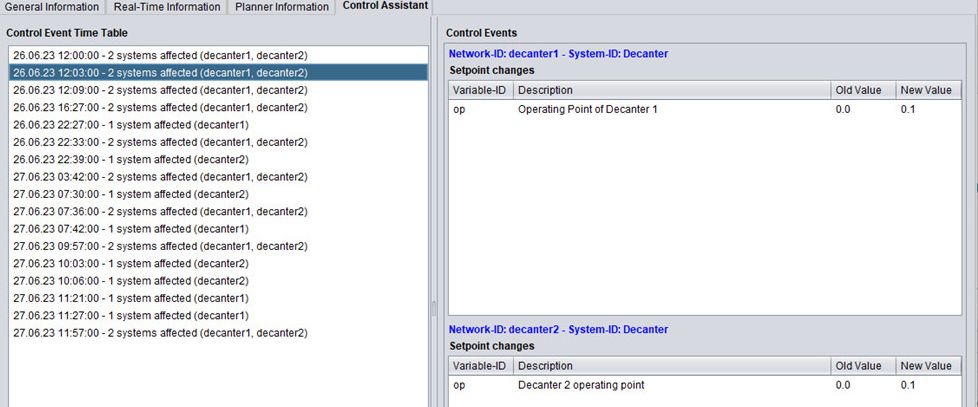}
    \caption{Assistenzsystem des \ac{eom}}
    \label{fig:eom:assi}
\end{figure}
\section{Optimierter, energieflexibler Betrieb der Abwasseraufbereitungsanlage} \label{sec:ergebnisse}
\noindent Dieser Abschnitt stellt die Ergebnisse der Anwendung des in \autoref{sec:method} beschriebenen Modells dar. Dazu wird zunächst in \autoref{sec:setup} der Setup des Feldtests beschrieben. In \autoref{sec:ergebnisse:results} werden die Ergebnisse der Umsetzung des optimierten Betriebsplans beschrieben. 

\subsection{Setup des Feldtests} \label{sec:setup}
Das in \autoref{sec:method:modell} vorgestellte Optimierungsmodell wurde mit den in \autoref{tab:parameter} gelisteten Parametern parametriert und unter Nutzung der Optimierungsumgebung des \ac{eom} auf einem Rechner mit Windows 11 und einem AMD Ryzen 7 PRO 5850U Prozessor mit 32 GB Arbeitsspeicher mit einer Optimalitätslücke von 10$^{-3}$ gelöst. Die \textit{Optimalitätslücke} ist ein Maß dafür, wie nah eine gefundene Lösung an der theoretisch optimalen Lösung eines \ac{milp} Problems liegt. Sie wird als das Verhältnis der Differenz zwischen dem besten gefundenen Lösungswert und einer unteren Schranke zum Lösungswert definiert und in Prozent ausgedrückt. Die zeitliche Auflösung des Optimierungsmodells ($\Delta t$) beträgt 3 Minuten.

Als Zielwert für die Optimierung wurde vorgegeben, dass der Füllstand der Schlammtasche am Ende des Optimierungshorizonts dem Startfüllstand von 350~m$^3$ entspricht. Zu Beginn befinden sich beide Dekanter im Zustand \texttt{Aus}.

Das Optimierungsmodell wurde in zwei Schritten gelöst.  Im ersten Schritt wurde ein Betriebsplan von 11:30 bis 00:00 des Folgetags berechnet. Nach Bekanntgabe der Strompreise des Folgetages um 15:00 wurde ein weiterer Fahrplan für 00:00 bis 14:00 des Folgetages unter Berücksichtigung der Endwerte des ersten Betriebsplans generiert. Die Berechnungsdauern betrugen 5,3 s (11:30 bis 00:00, Tag 1) und 6,1 s (00:00 bis 14:00, Tag 2). Den Betreibern wurden dann im Assistenzsystem (\autoref{fig:eom:assi}) Handlungsempfehlungen angezeigt, die diese dann in Steuerungsbefehle umsetzen.

\subsection{Ergebnisse des Feldtests} \label{sec:ergebnisse:results}
In \autoref{sec:ergebnisse:results:controlable} werden  die Ergebnisse der Umsetzung des Betriebsplans durch beide Dekanter interpretiert. Folgend werden in \autoref{sec:ergebnisse:results:stor} Wechselwirkungen des Dekanterbetriebs mit vor- und nachgelagerten Speichern  betrachtet. In \autoref{sec:bewertung} werden die Ergebnisse des flexiblen Betriebs mit einer statischen Betriebsweise verglichen.

\subsubsection{Steuerbare flexible Anlagen} \label{sec:ergebnisse:results:controlable}
\autoref{fig:optschedule} zeigt den optimierten Betriebsplan für beide Dekanter im Zeitfenster von 11:30 bis 14:00 am Folgetag sowie den jeweils aktuellen Strompreis des Intraday-Markts. 

\begin{figure}[h]
    \centering   
     \includegraphics[width=.75\linewidth]{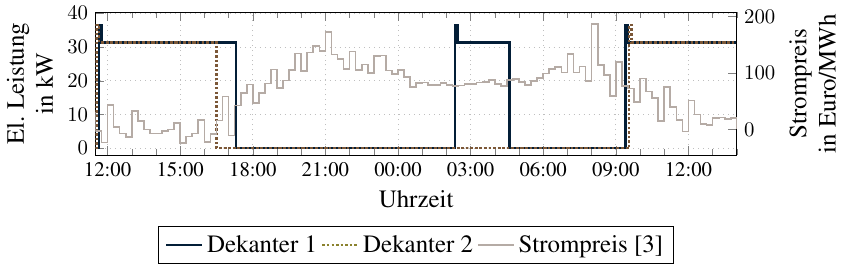}
    %\resizebox{.75\linewidth}{!}{\input{Abbildungen/opt_sched_dec1-dec2}}
\caption{Optimierter Betriebsplan zur Steuerung der Abwasseraufbereitungsanlage}
    \label{fig:optschedule}
\end{figure}

Darin zu sehen ist, gemäß der Zielfunktion (\autoref{eq:obj}) und unter Einhaltung der Nebenbedingungen aus \autoref{sec:method:modell}, die Planung der Nutzung elektrischer Energie und daher der  Betrieb der Dekanter in Zeiten mit verhältnismäßig geringen Strompreisen (z.B. nachmittags am ersten Tag). Ebenso ersichtlich ist dort das Einhalten von Haltedauern, um häufiges An- und Abschalten der Dekanter zu vermeiden. Der leichte Versatz der Betriebs\-pläne beider Dekanter, bspw. bei ca. 11:30 (Tag 1) oder ca. 09:30, ist durch das Verbot des gleichzeitigen \texttt{Starts} der Dekanter begründet. 
Der in \autoref{fig:optschedule} gezeigte Fahrplan wurde mit den Dekantern realisiert. 

\autoref{fig:comp:dec1} zeigt den Vergleich des Betriebsplans mit den im gleichen Zeitraum aufgezeichneten Messwerten. Darin zu sehen ist eine gute Übereinstimmung von Planung und Umsetzung, für die Variablen \glqq Elektrische Leistung\grqq \ (\autoref{fig:schedule:dec1:el}/\ref{fig:schedule:dec2:el}),  \glqq Dünnschlamm\grqq \ (\autoref{fig:schedule:dec1:sludge}/\ref{fig:schedule:dec2:sludge}) sowie  \glqq Trockenschlamm\grqq  \ (\autoref{fig:schedule:dec1:drysludge}/\ref{fig:schedule:dec2:drysludge}). 

\begin{figure}[h]
    \centering
        \subfloat[El. Leistung (Dek. 1) \label{fig:schedule:dec1:el}]{\includegraphics[height=3cm]{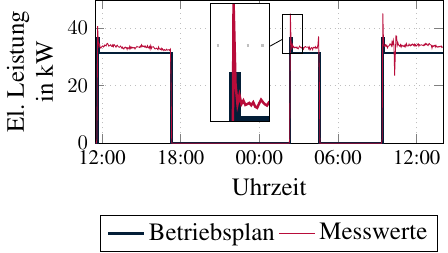}}
    \subfloat[Dünnschlamm  (Dek. 1)  \label{fig:schedule:dec1:sludge}]{ \includegraphics[height=3cm]{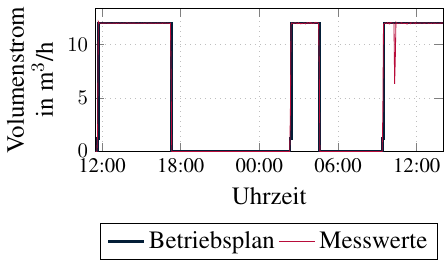}}
    \subfloat[Trockenschlamm  (Dek. 1) \label{fig:schedule:dec1:drysludge}]{ \includegraphics[height=3cm]{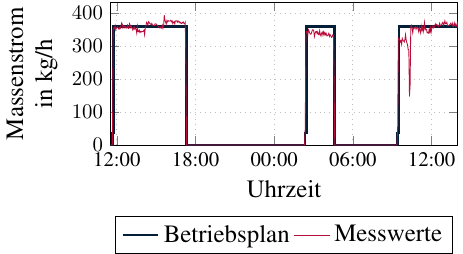}}  \\
    \subfloat[El. Leistung  (Dek. 2) \label{fig:schedule:dec2:el}]{ \includegraphics[height=3cm]{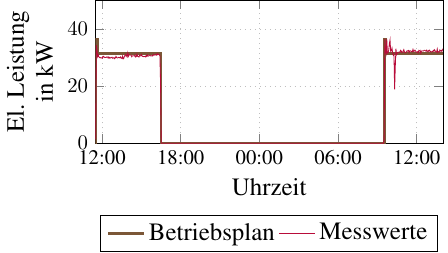}}
    \subfloat[Dünnschlamm  (Dek. 2) \label{fig:schedule:dec2:sludge}]{ \includegraphics[height=3cm]{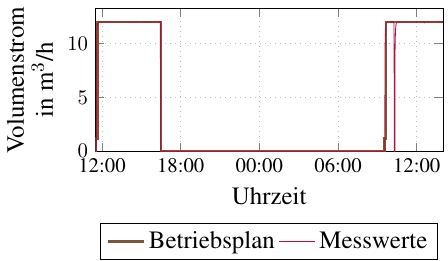}}
    \subfloat[Trockenschlamm  (Dek. 2) \label{fig:schedule:dec2:drysludge}]{ \includegraphics[height=3cm]{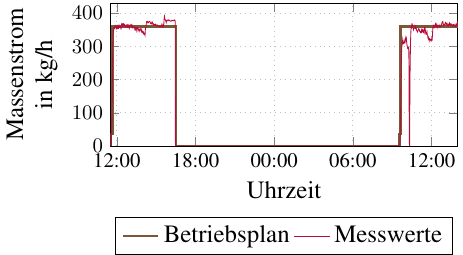}}
%        \subfloat[El. Leistung (Dek. 1) \label{fig:schedule:dec1:el}]{\resizebox{!}{3cm}{\input{Abbildungen/comp_sched_dec1_el}}}
%    \subfloat[Dünnschlamm  (Dek. 1)  \label{fig:schedule:dec1:sludge}]{\resizebox{!}{3cm}{\input{Abbildungen/comp_sched_dec1_duennschlamm}}}
%    \subfloat[Trockenschlamm  (Dek. 1) \label{fig:schedule:dec1:drysludge}]{\resizebox{!}{3cm}{\input{Abbildungen/comp_sched_dec1_trockenschlammfracht}}}  \\
%    \subfloat[El. Leistung  (Dek. 2) \label{fig:schedule:dec2:el}]{\resizebox{!}{3cm}{\input{Abbildungen/comp_sched_dec2_el}}}
%    \subfloat[Dünnschlamm  (Dek. 2) \label{fig:schedule:dec2:sludge}]{\resizebox{!}{3cm}{\input{Abbildungen/comp_sched_dec2_duennschlamm}}}
%    \subfloat[Trockenschlamm  (Dek. 2) \label{fig:schedule:dec2:drysludge}]{\resizebox{!}{3cm}{\input{Abbildungen/comp_sched_dec2_trockenschlammfracht}}}
    \caption{Vergleich einzelner Variablen des Betriebsplans der Dekanter mit Messwerten}
    \label{fig:comp:dec1}
\end{figure}

Folgende Interpretationen treffen auf beide Dekanter zu: %die Modelle

Die gemessene elektrische Leistungsaufnahme schwankt leicht, jedoch folgt sie grundsätzlich dem Wert der Planung. Die Startvorgänge führen zu Lastspitzen, die zwar im Modell abgebildet sind, jedoch nicht in ihrer konkreten Höhe (siehe Hervorhebung in \autoref{fig:schedule:dec1:el}). Durch die kurze Dauer ist diese Abweichung vernachlässigbar, unterstreicht jedoch die Relevanz der Nebenbedingung zur Vermeidung des gleichzeitigen Start-Vorgangs. Die gute Übereinstimmung der Leistungswerte kann durch die \ac{nrmse} von 4,5~\% (Dekanter 1) und 2,8~\% (Dekanter 2) quantifiziert werden.

Beim Dünnschlamm-Volumenstrom ist eine gute Übereinstimmung von Modell und Realität erkennbar (7,1~\%, Dekanter 1; 4,8~\%, Dekanter 2). Dies ist durch einen zuvor definierten, konstanten Volumenstrom von 12 m$^3$/h bedingt (Parameter~$e$, \autoref{tab:parameter}). Lediglich beim Ein- und Ausschalten kommt es zu kleineren Abweichungen, da das Modell diskrete Zustandsänderungen annimmt, die in der Realität jedoch kontinuierlich ablaufen. Da diese Zustandsänderungen bei jedem Ein- und Ausschalten der Dekanter unterschiedlich lange dauern, sind sie schwierig zu modellieren. Die Auswirkungen dieser Modellierungsungenauigkeiten wird im weiteren Verlauf diskutiert.

Der Trockenschlamm-Massenstrom (sowohl im Modell als auch im Prozessleitsystem) berechnet sich als Produkt des Dünnschlamm-Volumenstroms und der Trockenschlamm-Dichte (siehe \autoref{eq:trockenschlamm}). Die Trockenschlamm-Dichte wurde im Modell mit 30 g/l angenommen, schwankt jedoch in der Realität geringfügig. Daher entstehen kleinere Abweichungen zwischen Betriebsplan und Messwerten, quantifiziert durch einen \ac{nrmse}, von 6,8~\% (Dekanter 1, \autoref{fig:schedule:dec1:drysludge}) und 4,8~\% (Dekanter 2, \autoref{fig:schedule:dec2:drysludge}).

Die Abweichungen der Messwerte vom Betriebsplan um ca. 10:00 des Folgetags bei allen Variablen in \autoref{fig:comp:dec1} sind auf eine kurze Unterbrechung des Dünnschlamm-Volumenstroms zurückzuführen. Dadurch bedingt sank die elektrische Leistungsaufnahme und die entwässerte Trockenschlamm-Menge.

Im Rahmen der Auswertung des Labortests im Projekt \textit{OptiFlex} wurde  der sog. \textit{Strategy Evaluation Plot} entwickelt, welcher die Planerfüllung durch eine Anlage über den Optimierungshorizont darstellt. Dabei wird die Abweichung des Integrals einer Variable des Betriebsplans über die Zeit vom Integral des korrespondieren Messwerts bestimmt. Eine negative Abweichung repräsentiert dabei eine Unterschreitung des Plans, eine positive Abweichung eine Überschreitung. \cite{RWF23}

In \autoref{fig:sep:dec1} sind  \textit{Strategy Evaluation Plots} für beide Dekanter dargestellt. Dort sind alle relevanten Variablen für die Steuerung (Leistung, \autoref{fig:sep:dec1:el}) und die Überwachung des Betriebs (Dünnschlamm-Volumenstrom, \autoref{fig:sep:dec1:sludgein} und Trockenschlamm-Massenstrom, \autoref{fig:sep:dec1:sludgeout}) mit ihrem jeweiligen Integral sowie dessen Abweichung zu den Messwerten dargestellt. 

\begin{figure}[h]
    \centering
    \subfloat[Energieverbrauch  \label{fig:sep:dec1:el}]{\includegraphics[height=4.75cm]{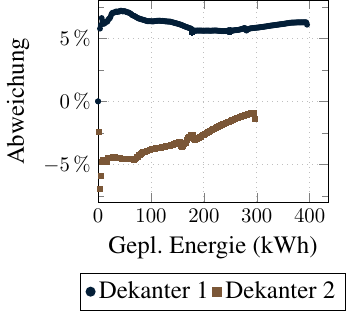}}
    \subfloat[Dünnschlamm  \label{fig:sep:dec1:sludgein}]{\includegraphics[height=4.75cm]{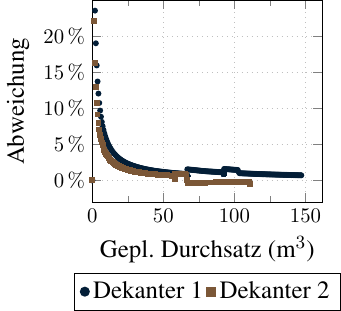}}
    \subfloat[Trockenschlamm  \label{fig:sep:dec1:sludgeout}]{\includegraphics[height=4.75cm]{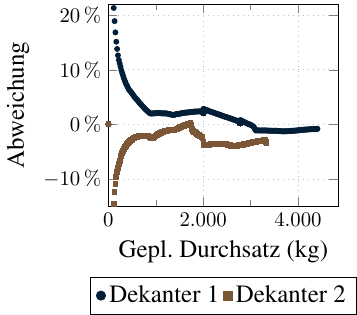}}
%    \subfloat[Energieverbrauch  \label{fig:sep:dec1:el}]{\resizebox{!}{4.75cm}{\input{Abbildungen/strat_eva_plot_dec1_el}}}
%    \subfloat[Dünnschlamm  \label{fig:sep:dec1:sludgein}]{\resizebox{!}{4.75cm}{\input{Abbildungen/strat_eva_plot_dec1_sludge}}}
%    \subfloat[Trockenschlamm  \label{fig:sep:dec1:sludgeout}]{\resizebox{!}{4.75cm}{\input{Abbildungen/strat_eva_plot_dec1_drysludge}}}
    \caption{Strategy Evaluation Plot}
    \label{fig:sep:dec1}
\end{figure}

Es ist ersichtlich, dass direkt zu Beginn des Optimierungshorizonts signifikante Abweichungen zwischen Plan und Messwerten entstehen. Dies ist bedingt dadurch, dass sich beide Dekanter zu Beginn des Optimierungshorizonts im Zustand \texttt{Aus} befinden. Durch die diskrete Modellierung der Betriebszustände führt der Start-Vorgang zu Abweichungen zwischen den beiden Integralen.   

Starke, quasi-sprunghafte Veränderungen bei den Abweichungen der Variablen \textit{Dünnschlamm} (\autoref{fig:sep:dec1:sludgein}) und \textit{Trockenschlamm} (\autoref{fig:sep:dec1:sludgeout}) innerhalb des Optimierungshorizonts sind ebenfalls auf die Startvorgänge zurückzuführen. 

Allgemein zeigt sich in \autoref{fig:sep:dec1} eine gute Realisierung des Betriebsplans, insbesondere da am Ende des Intervalls die Abweichung des Dünnschlamm- und Trocken\-schlammdurchsatzes nahe 0 ist. Dies bedeutet, dass der Betriebsplan bilanziell präzise umgesetzt wurde. Die Produktionsziele wurden somit erfüllt, wodurch keine negativen Auswirkungen auf den Betrieb zu erwarten sind. 

Es zeigt sich auch, dass die Übereinstimmung des Modells von Dekanter 1 mit dem Anlagenverhalten schlechter ist als die von Dekanter 2. Dies zeigt sich zum einen durch den höheren \ac{nrmse} sowie die verbleibende Abweichung am Ende des Optimierungshorizonts von 6~\% (siehe \autoref{fig:sep:dec1:el}). Diese Abweichungen können nach Rücksprache mit dem Anlagenbetreiber auf starke Verschmutzungen im Dekanter zurückgeführt werden. 

\subsubsection{Massespeicher} \label{sec:ergebnisse:results:stor}
Der Füllstand der Schlammtasche ist abhängig vom Dünnschlammzustrom und dem Betrieb der Dekanter (siehe \autoref{sec:method:modell}). In Ermangelung von Vorhersagen über die Menge des Zustroms wurde dieser im Modell als konstant definiert.  
\autoref{fig:comp:speicher} zeigt den Vergleich des Füllstands der Schlammtasche (Variable des Betriebsplans) mit den Messwerten.  Der Füllstand wird nur zu bestimmten Zeitpunkten (alle 8 Stunden) händisch erfasst. Eine kontinuierliche Aufzeichnung von Messwerten erfolgt nicht.  Daher ist in \autoref{fig:comp:speicher} auch der Messwert um 16 Uhr abgebildet, welcher außerhalb des Optimierungshorizonts liegt. 

\begin{figure}[h]
    \centering
\includegraphics[height=4cm]{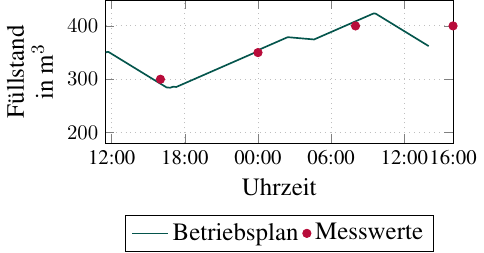}
     %\resizebox{!}{4cm}{\input{Abbildungen/comp_sched_mudbag_soc}}
    \caption{Vergleich des berechneten Füllstands der Schlammtasche mit Messwerten}
    \label{fig:comp:speicher}
\end{figure}

Ein qualitativer Vergleich der Messwerte mit dem Betriebsplan zeigt eine hinreichend gute Übereinstimmung, insbesondere vor dem Hintergrund des in der Realität variablen, und schwer vorherzusagenden Zustroms in die Schlammtasche.

Durch die geringe Trockenschlammdichte im Zeitraum des Feldtests wurden die Trocken\-schlammcontainer nicht vollständig gefüllt und daher nicht ausgetauscht. Eine Registrierung des Gewichts erfolgt jedoch nur beim Austausch. Daher liegen keine Informationen über die in den Containern enthaltene Trockenschlammmenge vor. Durch die gute Übereinstimmung von Betriebsplan und Messwerten für die Trockenschlammproduktion der Dekanter (\autoref{fig:schedule:dec1:drysludge}/\ref{fig:schedule:dec2:drysludge}) liegt die Vermutung nahe, dass auch hier die durch die Optimierung berechneten Werte gut mit der Realität übereinstimmen.

\subsubsection{Bewertung des flexiblen Betriebs} \label{sec:bewertung}
Einer flexiblen Betriebsweise dynamischen Strompreisen folgend wird ein finanzielles Einsparungspotential beigemessen \cite{RWK+23}.  Ein Vergleich des  energetisch flexiblen Betrieb des Anlagenverbunds nach dem in \autoref{fig:optschedule} gezeigten Betriebsplan mit einer statischen Betriebsweise mit einer konstanten Leistungsaufnahme von 26,09~kW (Mittelwert der Leistungsaufnahme beider Dekanter im flexiblen Betrieb) zeigt ein Einsparungspotential von 56~\% auf. Im flexiblen Betrieb fallen über den Betriebszeitraum von etwa 26 Stunden Energiekosten von 18,57~€ an, während der statische, also inflexible Betrieb - mit den Preisen aus \autoref{fig:strompreis-epex-feldtest} - Energiekosten von 42,28~€ verursacht. 

Der in diesem Beitrag dokumentierte Energie-flexible Betrieb der Abwasseraufbereitungsanlage wurde ohne jegliche zusätzlichen Investitionen in Hardware oder Software beim Anlagenbetreiber durchgeführt. Um die Anlage dauerhaft Energie-flexibel zu Betreiben, wären jedoch einige Investitionen vorteilhaft. So sollte Sensorik nachgerüstet werden, die den Füllstand der Schlammtasche dauerhaft erfasst, um Re-Optimierungen anstoßen zu können und so ein Über- oder Leerlaufen des Beckens zuverlässig auszuschließen. Für einen höheren Automatisierungsgrad bei der Umsetzung Energie-flexibler Betriebspläne müsste zudem eine Schnittstelle zwischen dem Assistenzsystem (\autoref{fig:eom:assi}) und dem Prozessleitsystem der Abwasseraufbereitungsanlage geschaffen werden, über die Steuerungssignale an die Anlage übermittel werden können. Auch der personelle Aufwand für die Erstellung von Optimierungsmodellen wäre in der Praxis nicht zu vernachlässigen. Die hier vorgestellten Hilfsmittel, die im Forschungsprojekt \textit{OptiFlex} entwickelt wurden, tragen dazu bei, diesen Aufwand zu verringern.
\section{Diskussion und Fazit} \label{4disconcl}
\noindent In diesem Beitrag wird die Erstellung und Nutzung von Optimierungsmodellen zur Optimierung des Betriebsplans energieflexibler, industrieller Anlagenverbünde untersucht. Ein Optimierungsmodell wurde in einer eigens für diesen Zweck geschaffenen Optimierungsanwendung erstellt und parametriert. Unter Einbezug aktueller Strompreise des kontinuierlichen Intraday-Markts erfolgte dann eine ökonomische Optimierung zur Minimierung der Energiekosten. 

Die Ergebnisse zeigen eine gute bis sehr gute Übereinstimmung von Modell und aufgezeichneten Messwerten. Größere Abweichungen zwischen dem optimierten Betriebsplan und den Messwerten zeigen sich nur bei dem Starten und Stoppen der Anlage. Weiterhin konnte durch das Optimierungsmodell festgestellt werden, dass die Effizienz eines der Dekanter sich aufgrund von Schmutzablagerungen verschlechtert hatte, was ebenfalls zu kleineren Abweichungen zwischen Betriebsplan und Messwerten führte. Insgesamt ist das Modell sehr gut geeignet, um den Betriebsplan der Abwasseraufbereitungsanlage zu optimieren. Dies zeigt sich durch die nahezu exakte Erfüllung der Betriebsziele in den gezeigten Strategy Evaluation Plots (\autoref{fig:sep:dec1}). Häufiges An- oder Abschalten der Anlage wird vermieden, der optimierte Betriebsplan stimmt gut mit dem realen Anlagenverhalten überein und konnte mit geringem Aufwand von den Betreibern umgesetzt werden. Die im Rahmen des Projekts OptiFlex erarbeitete Optimierungsumgebung (\autoref{fig:optimierungsumgebung}) in Kombination mit dem Energieoptionsmodell der EnFlexIT GmbH ermöglichen hier sowohl die vereinfachte Definition von Optimierungsmodellen als auch die übersichtliche Bereitstellung von Betriebsplänen für Betreiber. 

Im Rahmen dieser Anwendung wurde der Betriebsplan den Anlagenbetreibern in Form von Handlungsempfehlungen zur Verfügung gestellt. Eine Alternative hierzu wäre, den Betriebsplan mittels einer automatisierten Schnittstelle direkt an das Prozessleitsystem des Anlagenverbundes zu übermitteln und somit automatisch umzusetzen. Aufgrund einiger schwer vorherzusagender Einflussgrößen wie dem anfallenden Abwasservolumen und dem Feststoffgehalt des Abwassers (Trockenschlammdichte) wird eine automatisierte Ausführung von Betriebsplänen seitens der Anlagenbetreiber kritisch gesehen. Spontane Reaktionen auf unvorhergesehene Ereignisse zur Sicherstellung des reibungslosen Betriebs haben hier Vorrang gegenüber der Minimierung von Energiekosten, sodass auch für die Zukunft Handlungsempfehlungen zur Einsparung von Energiekosten als die beste Alternative gesehen werden.

In zukünftigen Anwendungen der im Projekt OptiFlex entwickelten Methoden und Werkzeuge könnte demonstriert werden, dass neben der Minimierung der Stromkosten auch andere Optimierungsziele wie die explizite Minimierung von Treibhausgasemissionen oder die Integration lokal erzeugter erneuerbarer Energie verfolgt werden können. Auch langfristige Veränderungen von Anlagen, wie die hier entdeckten Verschmutzungen in einem der Dekanter, können mit den entwickelten Methoden und Werkzeugen erkannt werden, was einen weiteren zukünftigen Anwendungsfall darstellt. 

\section*{Danksagung}
\noindent Diese Forschungsarbeit aus dem Projekt OptiFlex wurde durch dtec.bw -- Zentrum für Digitalisierungs- und Technologieforschung der Bundeswehr gefördert. dtec.bw wird von der Europäischen Union -- NextGenerationEU finanziert.

Die Autoren danken den Mitarbeitern der Abwasseraufbereitungsanlage für ihre Kooperation und Unterstützung im Rahmen des Feldtests.

\section*{Referenzen}
\printbibliography[heading=none]
% \bibliographystyle{model1-num-names} 
% \bibliographystyle{elsarticle-num-names}
% \bibliography{literature-optiflex}
\fontsize{10pt}{12pt}\selectfont

\section*{Autoren}
\noindent M.Eng.  Lukas Wagner (geb. 1995) ist seit November 2021 wissenschaftlicher Mitarbeiter am Institut für Automatisierungstechnik  der Helmut-Schmidt-Universität Hamburg. Seine Forschungsschwerpunkte sind die Modellierung energetischer flexibler Anlagen und das Energiemanagement im industriellen Kontext. \\

\noindent M.Eng. Lukas Wagner\\
Holstenhofweg 85\\
22043 Hamburg \\
\color{blue}
\underline{\href{lukas.wagner@hsu-hh.de}{lukas.wagner@hsu-hh.de}}
\color{black}\\

\noindent  M.Sc. Lasse Reinpold (geb. 1993) ist seit September 2021 wissenschaftlicher Mitarbeiter am Institut für Automatisierungstechnik der Helmut-Schmidt-Universität Hamburg. Seine Forschungsschwerpunkte liegen in der Automatisierung von energieoptimierter Produktionsplanung. \\

\noindent M.Sc. Maximilan Kilthau (geb. 1998)  ist seit April 2021 wissenschaftlicher Mitarbeiter am Institut für Automatisierungstechnik  der Helmut-Schmidt-Universität Hamburg. Seine Forschungsschwerpunkte sind die dezentrale, agentenbasierte Frequenz- und Spannungsregelung im elektrischen Verteilnetz. \\

\noindent Dr.-Ing. Felix Gehlhoff (geb. 1988) promovierte 2023 an der Helmut-Schmidt-Universität über agentenbasierte Steuerungsverfahren zur Optimierung des Produktions- und Logistikschedulings. Er war Forschungsgruppenleiter für agentenbasierte Systeme und anschließend Post-Doc für autonome Systeme und KI. Als Leiter der Abteilung Automatisierungstechnik umfassen seine aktuellen Forschungsinteressen Engineering-Methoden für komplexe Automatisierungssysteme, die Anwendung autonomer Systeme in Produktion, Energie und Intralogistik sowie die modellbasierte Darstellung, Steuerung und Validierung automatisierter Prozesse. \\

\noindent Dipl.-Ing. Christian Derksen (geb. 1970) ist seit Oktober 2022 Geschäftsführer der EnFlex.IT GmbH. Dort führt er die Entwicklung der Agent.Workbench sowie des Energieoptionsmodells fort, welche er im September 2009 als wissenschaftlicher Mitarbeiter am DAWIS/ICB der Universität Duisburg-Essen begann. Wissenschaftlich beschäftigt er sich mit adaptiven Regelungsansätzen im Bereich hybrider Energiesysteme.\\

\noindent M.Sc. Nils Loose (geb. 1981) ist seit Oktober 2022 Software-Entwickler für die Agent.Workbench und das Energieoptionsmodell bei der EnFlex.IT GmbH. Zuvor war er wissenschaftlicher Mitarbeiter am Lehrstuhl für Datenverwaltungssysteme und Wissensrepräsentation an der Universität Duisburg-Essen. Seinen Forschungsschwerpunkt bildet die Anwendung von Agentensystemen zur Steuerung und Optimierung intelligenter Energienetze.\\

\noindent Prof. Dr.-Ing. Julian Jepsen ist Professor für angewandte Werkstofftechnik an der Helmut-Schmidt-Universität Hamburg und leitet die Abteilung für Systemdesign für mobile Speicher am Helmholtz-Zentrum Hereon. Seine Forschung konzentriert sich auf die Entwicklung und Integration von wasserstoffbasierten Energiespeichersystemen, insbesondere solchen, die auf Metallhydriden basieren. Ein besonderer Schwerpunkt liegt dabei auf der Simulation dieser Systeme mittels digitaler Zwillinge und der Sektorkopplung zur Integration erneuerbarer Energien. \\

\noindent Prof. Dr.-Ing. Alexander Fay (geb. 1970) ist Professor für Automatisierungstechnik an der  Ruhr-Universität Bochum. Seine Forschungsschwerpunkte sind Beschreibungsmittel, Methoden und Werkzeuge für ein effizientes Engineering von Automatisierungssystemen und deren Betrieb. 
\end{addmargin}
\end{document}